\documentclass[a4paper,twocolumn,aps,prl]{revtex4-1}

\newcommand{\eq}[1]{\begin{equation}#1\end{equation}}
\newcommand{\dd}{\mathrm{d}}
\newcommand{\ee}{\mathrm{e}}

\newcommand{\Ai}{\mathrm{Ai}}

\newcommand{\Tr}{\mathrm{Tr \,}}
\usepackage{amsmath}
\usepackage{graphics}
\usepackage{graphicx}
\usepackage{psfrag}

\begin{document}

\title{Universality in the full counting statistics of trapped fermions}

\author{Viktor Eisler}
\affiliation{Vienna Center for Quantum Science and Technology,
Faculty of Physics, University of Vienna,
Boltzmanngasse 5, A-1090 Wien, Austria}

\date{\today}

\begin{abstract}
We study the distribution of particle number in extended subsystems of a one-dimensional
non-interacting Fermi gas confined in a potential well at zero temperature. Universal features are identified in the scaled
bulk and edge regions of the trapped gas where the full counting statistics are given by the corresponding limits
of the eigenvalue statistics in Gaussian unitary random matrix ensembles.
The universal limiting behavior is confirmed by the bulk and edge scaling of the particle number fluctuations
and the entanglement entropy.
\end{abstract}

\maketitle

The techniques for the manipulation of ultracold atomic gases have undergone a
rapid development in the last decade and have provided experimental access to
various interesting aspects of many-body quantum systems \cite{IKS07,BDZ08}.
The common feature in the experiments is the presence of trapping potentials
that can be tuned to confine particles into effective one-dimensional geometries
\cite{KWW04,Paredes04,Stoeferle04} and theoretical predictions on various properties
of 1D quantum gases \cite{GPS08,CCGOR11,GBL13} can directly be tested.

The strongly correlated phases of 1D quantum gases are characterized
by the simultaneous presence of thermal and quantum noise.
At  ultra low temperatures, the dominating quantum noise reveals important information
on the non-local character of correlations in the corresponding many-body states.
In particular, 1D Bose liquids can be probed by measuring the full distribution of interference
amplitude in experiments \cite{Hofferberth08}, showing remarkable agreement with the predictions of
the theory \cite{GADP06}. In case of a Fermi gas, an analogous concept is the
full counting statistics (FCS) \cite{Levitov} which encodes the distribution of
particle number in extended subsystems. The FCS shows interesting properties
in the ground state of the Fermi gas and, in the non-interacting case, can also be used
to extract the entanglement entropy of the subsystem \cite{Song12,CMV12}.

The presence of trapping potentials leaves characteristic signatures on the FCS.
For the ground state of the non-interacting Fermi gas, the FCS was studied in the presence of
a periodic potential \cite{BM04} and recently the effect of harmonic traps has been
analyzed on the particle number fluctuations and entanglement \cite{Vicari12a}.
However, the question whether some properties of the FCS hold irrespectively of
the details of the potential has not yet been addressed.

Here we point out a remarkable universality and show that, for a broad class
of trapping potentials, the proper scaling limits of the FCS in the bulk and edge regime of
the trapped gas are given by the corresponding eigenvalue statistics
of Gaussian unitary random matrix ensembles. Physically, the universality can be understood
from the generic behaviour of the trapping potential, being flat in the center and approximately
linear around the edge of the high-density region. The FCS is derived by a semi-classical treatment
of the single-particle wavefunctions and by finding scaling variables for the bulk and edge regimes
through which the details of the potential can be completely eliminated in the thermodynamical limit.

The appearance of random matrix eigenvalue statistics in the FCS is rooted in the free fermion nature of the problem.
However,  using the Fermi-Bose mapping \cite{Girardeau60} the results immediately carry over to the
bosonic Tonks-Girardeau gas. Since the latter one is accessible in cold-atom experiments
\cite{KWW04,Paredes04}, the measurement of the FCS might be feasible in the spirit of Ref. \cite{Hofferberth08}
where the full distribution function of an analogous observable could be extracted for trapped bosons.

The FCS is defined through the generating function
\eq{
\chi(\lambda) = \langle \exp ( i \lambda N_A )\rangle ,
\quad N_A = \int_{A} \rho(x) \dd x
}
where $N_A$ is the total number of particles in subsystem $A$,
given by the integral of the density $\rho(x)$, and the expectation value
is taken with the $N$-particle ground state of the system.
For the spinless free Fermi gas, the FCS can be expressed as a Fredholm determinant
\cite{CMV12,IAC11}
\eq{
\chi(\lambda) =  \det \left[1+(\ee^{i\lambda}-1) K_A \right]
\label{eq:chi}
}
where $K_A$ is an integral operator with the kernel
\eq{
K_A(x,y) = \sum_{k=0}^{N-1} \varphi^{*}_k(x) \varphi_k(y)
\label{eq:kxy}}
given by the two-point correlation function restricted to the domain $x,y \in A$.
It is constructed from the single-particle eigenfunctions of the Schr\"odinger equation
\eq{
\frac{1}{2} \frac{\dd^2 \varphi_k(x)}{\dd x^2} + \left[ E_k-V(x) \right] \varphi_k(x) = 0
\label{eq:seq}}
where we we have set $\hbar=m=1$.
For simplicity, we consider a symmetric $V(-x)=V(x)$ and monotonously increasing
trapping potential such that for $x \to \infty$ one has $V(x)\to\infty$. 
The spectrum is thus discrete and for each $k$ the solutions $\varphi_k(x)$ admit
two classical turning points given by the condition $V(\pm x_{0k}) = E_k$. In the
following we will consider $x \ge 0$ since the real valued wavefunctions must obey
the symmetry $\varphi_k(-x)=(-1)^k \varphi_k(x)$.

In the classically allowed region, $x<x_{0k}$, the wavefunctions are oscillatory with
exactly $k$ nodes whereas for $x>x_{0k}$ they must vanish exponentially. The wavefunction
which approximates the exact solution on both sides is known as the \emph{uniform Airy approximation}
and can be derived from a semi-classical treatment of the Schr\"odinger equation \cite{Airy}.
Up to the normalization factor $C_k$, it is given by
\eq{
\psi_k(x) = \frac{C_k}{\sqrt{\xi'_k(x)}}\Ai \left[ \pm \xi_k(x)\right]
\label{eq:unai}}
where the $+$ ($-$) sign applies in the classically forbidden (allowed) region
and the argument of the Airy function is given through
\eq{
\xi_k(x) = \left[ \frac{3}{2}\int_{x_1}^{x_2} p_k(z) \dd z \right]^{2/3}
\label{eq:xix}}
with $x_1 = \min (x,x_{0k})$ and $x_2 = \max (x,x_{0k})$.
The momentum in the integrand of Eq. (\ref{eq:xix}) is defined as $p_k(x)=\sqrt{2|E_k-V(x)|}$ and
the approximate energy niveaus $E_k$ can be obtained from the Bohr-Sommerfeld quantization
formula \cite{LL}
\eq{
\int_{-x_{0k}}^{x_{0k}} p_k(z) \dd z = (k+1/2) \pi .
\label{eq:bsq}}

In particular, we will be interested in power-law potentials of the form $V_p(x)=x^p/p$
with some even integer $p$. Note, that we have set the characteristic length scale of the
trap to one, which can easily be restored using the arguments of trap size scaling
\cite{CV10b,*CV10c}. Then the integral in Eq. (\ref{eq:bsq}) yields
\eq{
E_k \approx \left[\mathcal{N}_p (k+1/2)\right]^{2\theta} \, , \quad
\mathcal{N}_p = \frac{\sqrt{\pi} \, \Gamma(3/2+1/p)}
{\sqrt{2} p^{1/p}\, \Gamma(1+1/p)} 
\label{eq:bsp}}
with the exponent given by $\theta=p/(p+2)$.

\begin{figure}[thb]
\center
\includegraphics[width=\columnwidth]{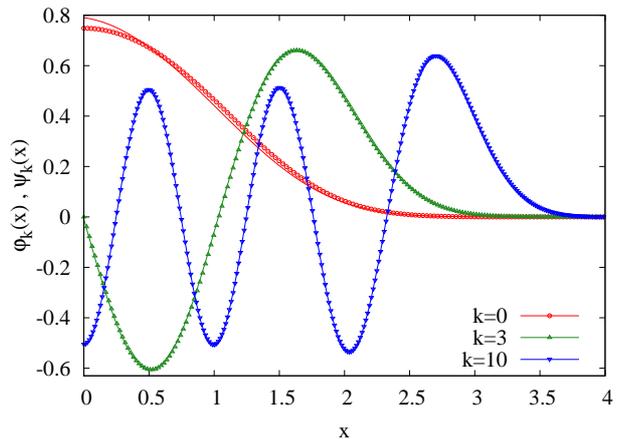}
\caption{(color online) Exact wavefunctions $\varphi_k(x)$ (symbols) and their uniform
Airy approximations $\psi_k(x)$ (lines) for the quartic potential $V_4(x)$.}
\label{fig:phipsi}
\end{figure}

In general, the approximate wavefunctions $\psi_k(x)$ given by Eq. (\ref{eq:unai})
rely on a semi-classical argument and are thus expected to reproduce the exact ones
$\varphi_k(x)$ only for $k \gg 1$. In fact, however, the uniform Airy approximation gives
very good results even for the eigenfunctions of the low-lying levels. This is demonstrated
on Fig. \ref{fig:phipsi} for the quartic potential $V_4(x)$. The eigenfunctions $\varphi_k(x)$ are calculated
to a high precision by Numerov's method \cite{Vesely} and compared to $\psi_k(x)$ where the integrals
in Eq. (\ref{eq:xix}) can be given through special (hypergeometric and incomplete beta) functions and
evaluated numerically. While the overlap is reasonable for $k=0$, the deviations are already very small
for $k=3$ and the two functions are essentially indistinguishable for $k=10$.

%
%

Far away from the turning point $x_{0k}$, the uniform Airy approximation (\ref{eq:unai}) reproduces
the WKB-approximation \cite{Airy} which can be obtained from asymptotic expansions of
$\Ai \left[\pm \xi_k(x)\right]$. In particular, in the classically allowed region one has
\eq{
\psi_k(x) = \frac{C_k}{\sqrt{\pi p_k(x)}}\cos \left(\int_{x}^{x_{0k}}p_k(z)\dd z -\frac{\pi}{4}\right) .
\label{eq:wkb}}
For simple potential wells with two classical turning points, the normalization constant can be
fixed by imposing \cite{Furry}
\eq{
C_k^2 \int_{-x_{0k}}^{x_{0k}} p^{-1}_k(z) \dd z = 2 \pi .
\label{eq:ck}}
Differentiating Eq. (\ref{eq:bsq}) with respect to $k$, one arrives to the simple formula
$C^2_k = 2 \frac{\dd E_k}{\dd k}$ and thus the normalization factor accounts for the spectral density.

The WKB-form of the wavefunctions (\ref{eq:wkb}) gives a good approximation in the entire classically
allowed regime not too close to the turning points, but depends on the details of the potential $V(x)$.
To find universal features of the FCS, we first focus on the \emph{bulk} of the trapped gas and we
choose the subsystem as the interval $A = \left[ -\ell,\ell\right]$ deep in the high-density region, $\ell \ll x_{0N}$.
Expanding the functions $p_k(x)$ around $x=0$, one obtains
\eq{
\psi_k(x) \approx \frac{C_k}{\sqrt{\pi \sqrt{2E_k}}} \cos \left( k\frac{\pi}{2} - \sqrt{2E_k}x\right)
\label{eq:wkbx0}}
which is valid up to linear terms in $x$.
Substituting into Eq. (\ref{eq:kxy}), one arrives to the following sum
%
\begin{align}
K_A(x,y) \approx
&\sum_{k=0}^{N-1} \frac{C_k^2}{2\pi \sqrt{2E_k}}
\cos\sqrt{2E_k}(x-y) \nonumber \\
+&\sum_{k=0}^{N-1} \frac{C_k^2(-1)^k}{2\pi \sqrt{2E_k}}
\cos\sqrt{2E_k}(x+y) \, .
\label{eq:kxybulk}
\end{align}
%
%
%
%

We are interested in the  $N \to \infty$ limit of the FCS, where the first sum in
Eq. ($\ref{eq:kxybulk}$) diverges. Therefore we introduce new variables and
define the scaling limit of the kernel as
\eq{
K_r(u,v) = \lim_{N\to\infty}\frac{1}{\sqrt{2E_N}}
K_A \left(\frac{u}{\sqrt{2E_N}},\frac{v}{\sqrt{2E_N}} \right)
\label{eq:ktuv}}
where the subscript refers to the domain $\left[ -r,r\right]$ of the kernel in the scaled
variables with the effective length $r=\ell \sqrt{2E_N}$. Introducing the variable
$z=\sqrt{E_k/E_N}$, the first sum in (\ref{eq:kxybulk}) can be converted
into an integral while the second, alternating sum vanishes in the scaling limit.
Writing $\dd z = \frac{\dd E_k}{\dd k}/\sqrt{4E_kE_N}$ and using the
expression of $C^2_k$ in terms of the spectral density, one finds
\eq{
K_r(u,v) = \frac{1}{\pi}\int_{0}^{1} \dd z \cos z (u-v) = \frac{\sin (u-v)}{\pi(u-v)} .
\label{eq:ksin}}
Hence, in the bulk scaling limit, we recover the sine kernel which appears in the theory of GUE random matrices.
Indeed, the probability $E(n,t)$ of finding $n$ eigenvalues in an interval $\left[-r,r\right]$ in the bulk
of the GUE spectrum is given by \cite{Mehta}
\eq{
E(n,r) = \left. \frac{(-1)^n}{n!} \frac{\dd^n}{\dd z^n} \det (1-z K_r) \right|_{z=1} .
\label{eq:enr}}
Fourier transforming Eq. (\ref{eq:enr}) with respect to $n$ yields the determinant in Eq. (\ref{eq:chi})
with $K_A=K_r$ and thus the bulk FCS of the Fermi gas is identical to the bulk GUE eigenvalue statistics.

The bulk scaling limit can be tested through the cumulants
$\kappa_m = \left. (-i \partial_\lambda)^m \ln \chi (\lambda) \right|_{\lambda=0}$ of the particle number.
In particular, we calculated the fluctuations $\kappa_2$ as a function of $\ell$ for the potential $V_4(x)$.
The numerics is simplified by considering,
instead of the integral operator $K_A$, the overlap matrix $\mathbf{C}_A$ with elements
\eq{
C_{A,kl} = \int_{A} \dd x \, \varphi_k(x) \varphi_l(x)
\label{eq:cmn}}
and using $\Tr K^n_A = \Tr \mathbf{C}^n_A$ \cite{CMV11a,*CMV11b}. Then the
FCS of Eq. (\ref{eq:chi}) can be treated as a regular determinant and one has
$\kappa_2(\ell) = \Tr \mathbf{C}_A (\mathbf{1}-\mathbf{C}_A )$. This is then compared
to existing results derived using the asymptotics of Fredholm determinants with the
sine kernel \cite{IAC11}
\eq{
\kappa_2(r) = \Tr K_r(1-K_r)=\frac{1}{\pi^2}(\log 4r + \gamma +1)
\label{eq:k2r}}
where $\gamma$ is the Euler constant and we neglected terms vanishing for $N\to\infty.$

\begin{figure}[thb]
\psfrag{L}[][][1]{$\ell$}
\center
\includegraphics[width=\columnwidth]{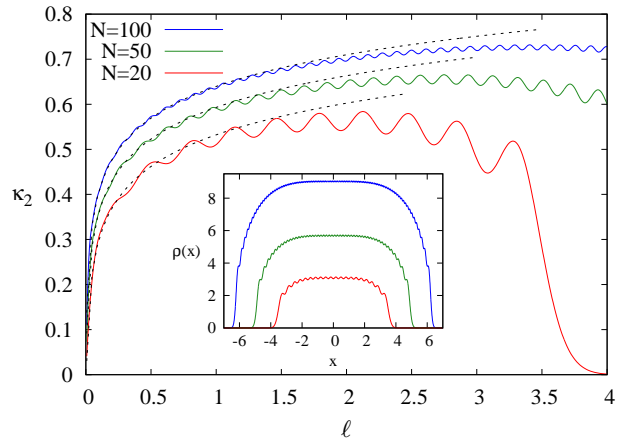}
\caption{(color online) Particle number fluctuations for the quartic potential $V_4(x)$ in the interval
$\left[-\ell,\ell\right]$ (solid lines) compared to the prediction of Eq. (\ref{eq:k2r}) in the bulk scaling
limit (dashed lines) for various $N$. The inset shows the corresponding density profiles $\rho(x)$.}
\label{fig:k2}
\end{figure}

The results are shown on Fig. $\ref{fig:k2}$ for various $N$ with the dashed lines
representing the curves $\kappa_2(r)$. One can see a good agreement for small $\ell$
but the solid curves $\kappa_2(\ell)$ deviate from the scaling prediction as soon as
the segment size exceeds the size of the flat region in the densities $\rho(x)=K_A(x,x)$,
shown on the inset. However, the amplitude of the $\mathcal{O}(x^2)$ corrections to
Eq. (\ref{eq:kxybulk}) is proportional to $V''(0)/E_k$ which vanishes for $V_p(x)$ with
$p \ge 4$ and thus the flat density region extends for higher $p$. Note also the strong
oscillations in $\kappa_2(\ell)$ as well as in $\rho(x)$ that are results of the alternating
sum in Eq. (\ref{eq:kxybulk}) and diminish for higher $N$. The bulk scaling
was further tested by calculating the entanglement entropy
%
$S(\ell) = - \Tr \left[ \mathbf{C}_A \ln \mathbf{C}_A + (1-\mathbf{C}_A) \ln (1-\mathbf{C}_A) \right]$
%
and comparing it to the scaling prediction $S(r)$ \cite{CMV11a, *CMV11b, SI12}
with similarly looking results as in Fig. \ref{fig:k2}.

%
%

The other regime where universal features are expected to emerge is near the edge of the
high-density region. Close to the classical turning point, the argument $\xi_k(x)$ of the Airy
function in (\ref{eq:unai}) can be expanded around $x_{0k}$ and yields \cite{Airy}
\eq{
\psi_k(x) \approx \frac{C_k}{\sqrt{\alpha_k}} \Ai \left[\alpha_k(x-x_{0k})\right]
\label{eq:aitp}}
with $\alpha_k = (2V'(x_{0k}))^{1/3}$ giving the inverse of the typical length scale.
The subsystem is now fixed as the interval $A=\left(x_{0N} + s/\alpha_N,\infty\right)$
starting close to the edge of the high-density region and extending to infinity.
Note, that this choice of the interval $A$ strongly limits the terms contributing to the sum in
Eq. (\ref{eq:kxy}) since the Airy functions in Eq. (\ref{eq:aitp}) are shifted gradually to the left for
decreasing $k$ and for $|x_{0k}-x_{0N}| \gg |s|/\alpha_N$ they become exponentially small in $A$.
The edge scaling limit of the kernel is then defined as
\eq{
K_{s}(u,v) = \lim_{N\to\infty} \frac{1}{\alpha_N}
K_A \left( x_{0N} + \frac{u}{\alpha_N},  x_{0N} + \frac{v}{\alpha_N}\right) 
\label{eq:ksuv}}
%
%
\begin{figure}[thb]
\center
\includegraphics[width=\columnwidth]{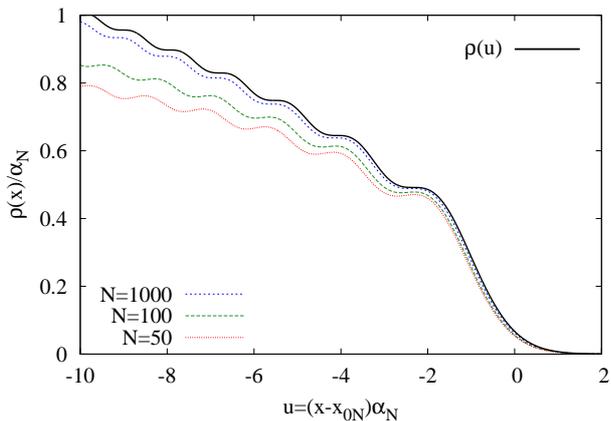}
\caption{(color online) Rescaled density profiles (dashed lines) for $V_4(x)$ near the edge
of the high-density region for various $N$.
The uppermost (solid) line shows the $N\to\infty$ scaling function $\rho(u)$, see text.}
\label{fig:dens}
\end{figure}
%
%
where the subscript refers to the domain $u,v \in \left(s,\infty\right)$ in the new variables.
The factors $\alpha_k/\alpha_N$ appearing in the arguments of the Airy functions can be
approximated as
\eq{
\frac{\alpha_k}{\alpha_N} \approx 1 + \frac{V''(x_{0N})}{3V'(x_{0N})} (x_{0k}-x_{0N})
\label{eq:akan}}
and since $|x_{0k}-x_{0N}| \sim |s|/\alpha_N$, the second term vanishes in the limit
$N \to \infty$ for any well behaved potential. We thus set $\alpha_k = \alpha_N$
in evaluating (\ref{eq:ksuv}) and introduce $z=\alpha_N(x_{0N}-x_{0k})$. Using $x_{0k}=V^{-1}(E_k)$,
one finds $\dd z \approx -C^2_k/\alpha^2_N$ and consequently
\begin{align}
K_s(u,v) &= \int_{0}^{\infty} \dd z \Ai(u+z) \Ai(v+z) \nonumber \\
&= \frac{\Ai(u)\Ai'(v)-\Ai'(u)\Ai(v)}{u-v} .
\label{eq:ksuv2}
\end{align}
Thus we recover the Airy kernel in the edge scaling limit. The FCS is then identical to the
GUE edge eigenvalue statistics \cite{TW94}, which follows immediately from a formula
analogous to (\ref{eq:enr}) by replacing $r$ with $s$.

As a first test of the scaling limit, we calculated the edge density profiles $\rho(x)$ in the potential
$V_4(x)$ for various $N$ and compared them to the density scaling function
$\rho(u)=K_s(u,u) = (\Ai'(u))^2 - u \Ai^2(u)$, with the result shown on Fig. \ref{fig:dens}. 
To reach larger particle numbers $N$, we used, instead of the exact wavefunctions $\varphi_k(x)$,
the $\psi_k(x)$ in Eq. (\ref{eq:unai}), that gives excellent results for the
profile in the edge region. The finite-size scaling of the data can be inferred from the correction
term in Eq. (\ref{eq:akan}). For power-law potentials one has $V''_p(x_{0N})/V'_p(x_{0N})=(p-1)/x_{0N}$
and multiplying by $(x_{0k}-x_{0N}) \sim \alpha^{-1}_N$, we obtain a $N^{-2/3}$ scaling of the finite-size corrections, which is
consistent with the data in Fig. \ref{fig:dens}. Note, that the exponent of $N$ is independent of $p$,
however, the prefactor $p-1$ implies that the scaling collapse gets worse for larger $p$ which
we indeed observed in the numerics for $p=6$.

%
\begin{figure}[thb]
\center
\includegraphics[width=\columnwidth]{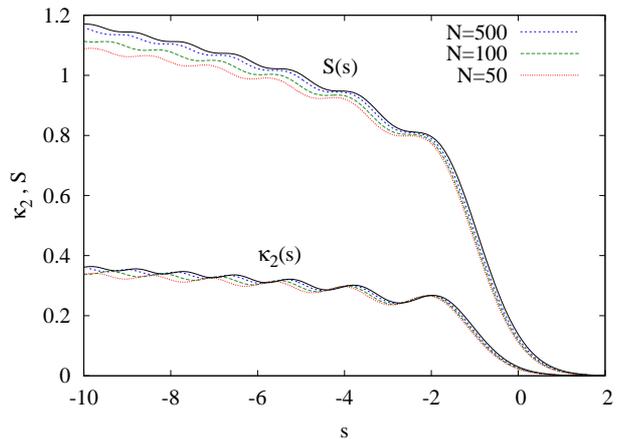}
\caption{(color online) Particle number fluctuations $\kappa_2$ and entropy $S$ in the scaled edge region
for various $N$. The solid lines show the respective scaling functions $\kappa_2(s)$ and $S(s)$.}
\label{fig:k2s}
\end{figure}
%

The edge limit of the FCS is verified by calculating the particle number fluctuations $\kappa_2$
and entanglement entropy $S$ of the interval $\left(s,\infty \right)$ in the scaled coordinates.
They are shown in Fig. \ref{fig:k2s} for various $N$ and compared to the scaling prediction,
calculated using a powerful numerical toolbox for the evaluation of Fredholm
determinants arising in random matrix theory \cite{Bornemann10}. Both $\kappa_2$ and $S$
converge slowly to their respective scaling functions.

In conclusion, we have shown the universality of the FCS for a noninteracting trapped Fermi gas
in the bulk and edge regions, given by the respective eigenvalue
statistics of GUE random matrices. Interestingly, the same universal limits emerge
for \emph{non-Gaussian} unitary random matrix ensembles, where the potential $V(x)$ appears in the
exponential weight function and the semiclassical asymptotics of the corresponding orthogonal
polynomials has to be analyzed \cite{BI99,DKMVX99}.
This leads to results that are surprisingly similar to Eq. (\ref{eq:unai}) even though the two problems
coincide only in the trivial Gaussian case, $V(x)$ being the harmonic potential.

One expects that the same universality would emerge for trapped fermions on a lattice and
might even generalize to other potentials.
In fact, the connection between FCS and GUE statistics has recently been pointed out
for the time evolution of lattice fermions from a step-initial condition \cite{ER13}.
However, in this case the correlation matrix is unitarily equivalent to the one describing the ground state
of a chain with a \emph{linear} potential \cite{EIP09} and hence the universality of the FCS carries over to
the gradient problem. Note that without the lattice, the Airy functions in (\ref{eq:aitp}) are the \emph{exact}
eigenstates and the spectrum is continuous, thus the edge limit (\ref{eq:ksuv2}) describes the FCS in
the entire high-density region while the bulk regime is missing.

It would also be interesting to consider the dynamical FCS after releasing the gas from the trap.
In the harmonic case, the corresponding time-dependent Schr\"odinger equation can be solved
exactly and the kernel has, up to an irrelevant phase factor, the equilibrium form in appropriately
rescaled variables \cite{Vicari12b}. Thus, the limiting scaling forms of the FCS are also unchanged.
However, for general $V(x)$ the situation is more complicated and requires a careful analysis.
Finally, in this non-equilibrium context one could study the waiting time distribution between
counting events where connections to random matrix theory have recently been revealed \cite{AHFB12}.

The author thanks Z. R\'acz for interesting discussions and acknowledges support by the ERC grant QUERG.



%

\end{document}